# SEISMIC PEAK PARTCILE VELOCITY AND ACCELERATION RESPONSE TO MINING FACES FIRING IN A LIGHT OF NUMERICAL MODELING AND UNDERGROUND MEASUREMENTS

**Witold Pytel**[1], **Piotr Mertuszka**[1], **Adam Lurka**[2], **Krzysztof Fuławka**[1], **Marcin Szumny**[1]

[1] KGHM CUPRUM Ltd. Research and Development Centre, **Poland**
[2] Central Mining Institute, **Poland**

## ABSTRACT

*Extraction of the copper ore deposit in the Legnica-Głogów Copper Basin in Poland is usually associated with high seismic activity. In order to face this threats, a number of organizational and technical prevention methods are utilized, from which blasting works seem to be the most effective. A significant number of recorded dynamic events may be clearly and directly explained by the effects of this approach. It is also expected, that the simultaneous firing of a number of mining faces may provide the amplification of vibrations in a specific location chosen within the rock mass. For better recognition of a such process, formation of an elastic wave generated by the detonation of explosives in a single mining face have been evaluated using the numerical tools and verified by the field measurements of ground particle velocity and acceleration parameters, i.e. PPV and PPA parameters. The primary objective of presented paper was to find the bridge between numerical simulations of the time-dependent seismic particle velocity values induced by blasting and in situ measurements using seismic three component geophones.*

## 1. INTRODUCTION

The three underground copper mines belonging to KGHM Polska Miedź S.A. are located in the Southwest part of Poland, Upper Silesia region. The flat copper ore exploitation is typically conducted through the use of blasting technology. Different types of mining methods based on room-and-pillar system have been developed as a result of many years' experience. In this system, the ore-seam is cut into passages and chambers separated by structural pillars. In the course of removal of rock the pillars degrade enabling gradual roof subsidence [1]. With the increasing depth of exploitation it was observed that increasing stress create more and more difficulties during the mining process [2]. Currently, the most significant hazard is associated with seismic activity resulting from disturbance of equilibrium state of primary stresses in the rock mass. Presence of stiff rocks in main roof strata results in strain energy accumulating . This in turn favours strong seismic events occurrence, when the energy is emitted into the surrounding rock mass in the form of elastic wave. The strongest mining tremors could reach the seismic energy over $10^9$ J and can be considered as small earthquakes, quite often associated with rockbursts at the excavations' level. Thus, high seismic activity may cause unfavourable impact on work safety. Some of the seismic events can generate the rockbursts that cover the working areas, what is related with dynamic character of rock mass failure. For better understanding of this type of seismicity, the stronger tremors' focal mechanisms are investigated on the routine bases in the local mines' seismic stations. Figure 1 shows the number of seismic events with the energy greater than $10^3$ J in the all KGHM mines, starting from 1990.

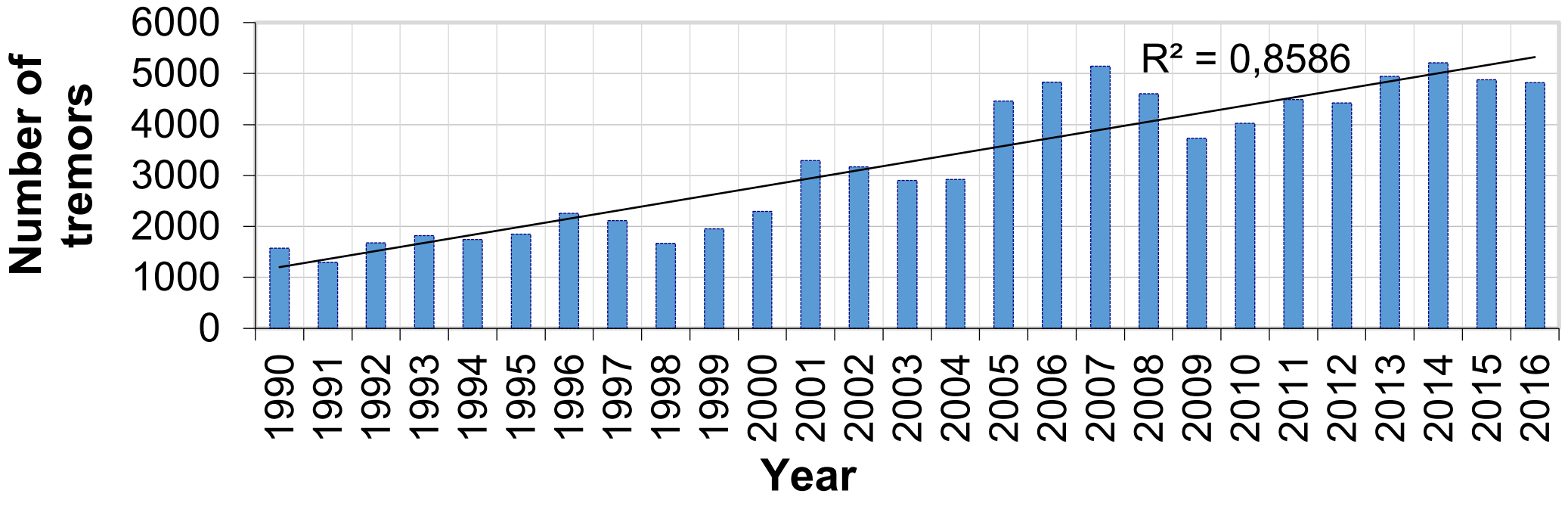

**Fig. 1. Number of tremors recorded between 1990 and 2016 in KGHM mines**

Based on Figure 1 one may conclude that the number of seismic events have generally increased in the past years. The general trend may be approximated by a linear function with the relatively high coefficient of determination.

Blasting works have been recognized to be an effective active method of seismic event prevention, conducted as multi-face strain-release production blasting [1]. This type of blasting works may increase the capability of inducing strain relief in the rock mass manifesting itself as  the induced seismic event occurring within the main roof strata. The detonation of explosives generates a propagating shock wave which may cause a serious damage to a material body that is encountered on its way. This wave in the form of mechanical vibrations is treated as the side effect, which has a destructive impact on the immediate surroundings [3, 4].

The authors' overall goal was to find the bridge between numerical simulations of the time-dependent distribution of seismic particle velocities induced by blasting and in situ measurements by seismic geophones.. Computer simulations of the propagation of an elastic wave generated by blasting operations were performed using a numerical model based on the geometry of the mine workings in the area of selected mining panel located within one of the Polish copper mines. The analyses have been supported by a 3D FEM modelling approach.

## 2. MATERIALS AND METHODS

For the purpose of analysis, G-4/8 mining panel located within G-23 mining division of the Rudna mine was selected. The deposit in this area is at a depth of 1,150 m. The following criteria were considered while selection of trial panel location: local degree of geomechanical hazard, existing mining infrastructure and speed of mining.

The abovementioned mining panel has been equipped with 16 geophones. It allowed to record the seismic wave beams generated by firing of explosives. Calibration of numerical model has been performed by verification of in situ three component measurements with calculated particle velocities. Records from two geophones were taken into account: 14 – located 155 m from the blasted mining face (MF) and 15 – located 140 m from the firing area (Fig. 2).

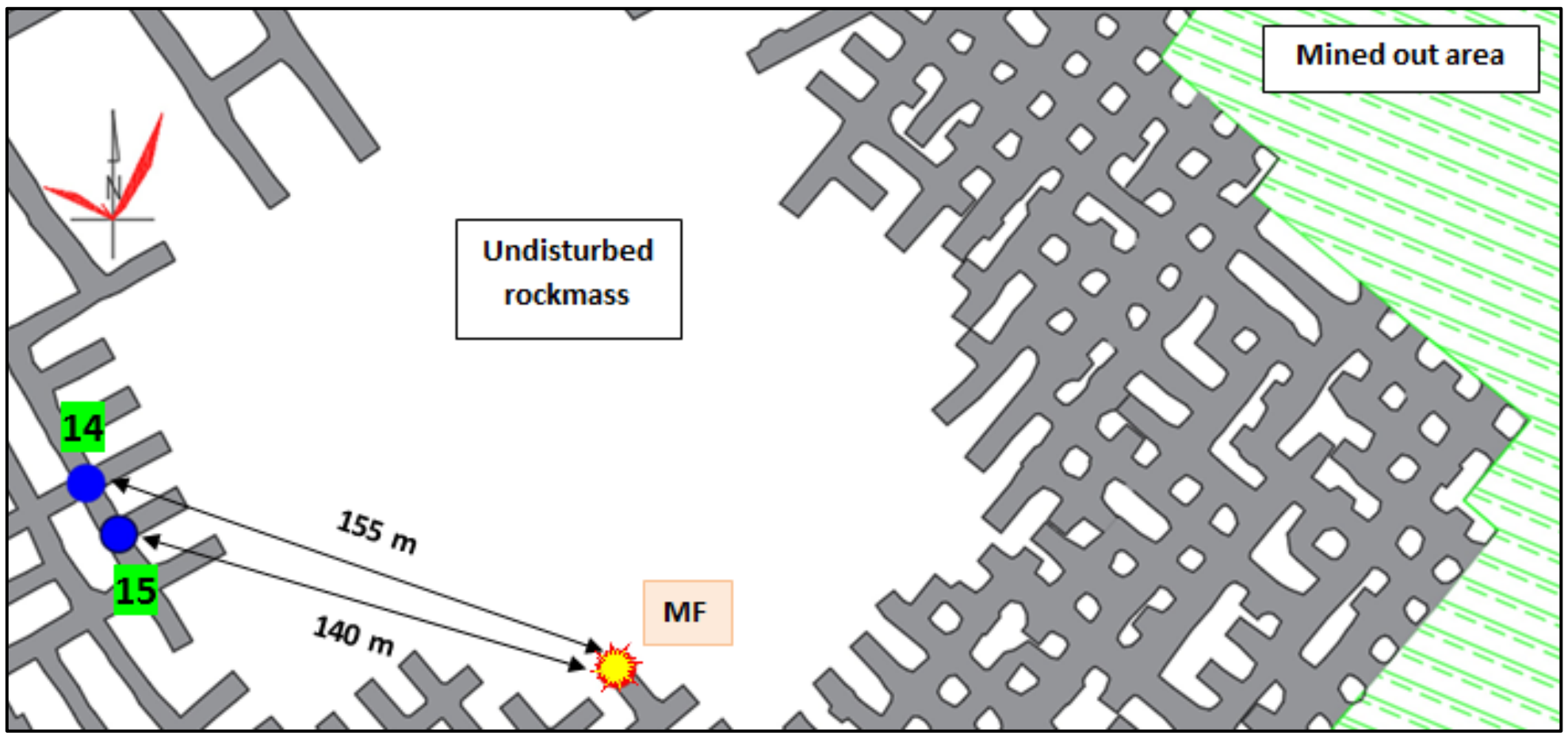

**Fig. 2. Location of geophones and blasted mining face**

The mining face firing pattern consisted of 20 blastholes, of 48 mm in diameter. The total amount of detonated explosives was 54 kg (2.7 kg per blasthole), 10.8 kg of which was charged into the cut holes. To illustrate the effect of mining face firing, the vicinity of G-4/8 mining panel has been modelled and analyzed using the finite element method formulated in three dimensions (Fig. 3). Geomechanical problem solution were based on the NEi/NASTRAN computer program code utilizing FEM in three dimensions [5]. It was assumed that the overburden strata consists of several homogeneous rock plates reflecting the real lithology in the area and the technological and remnant pillars work effectively within a post-critical phase (elastic-plastic with strain softening behaviour) [6]. A mining system with roof deflection has been assumed.

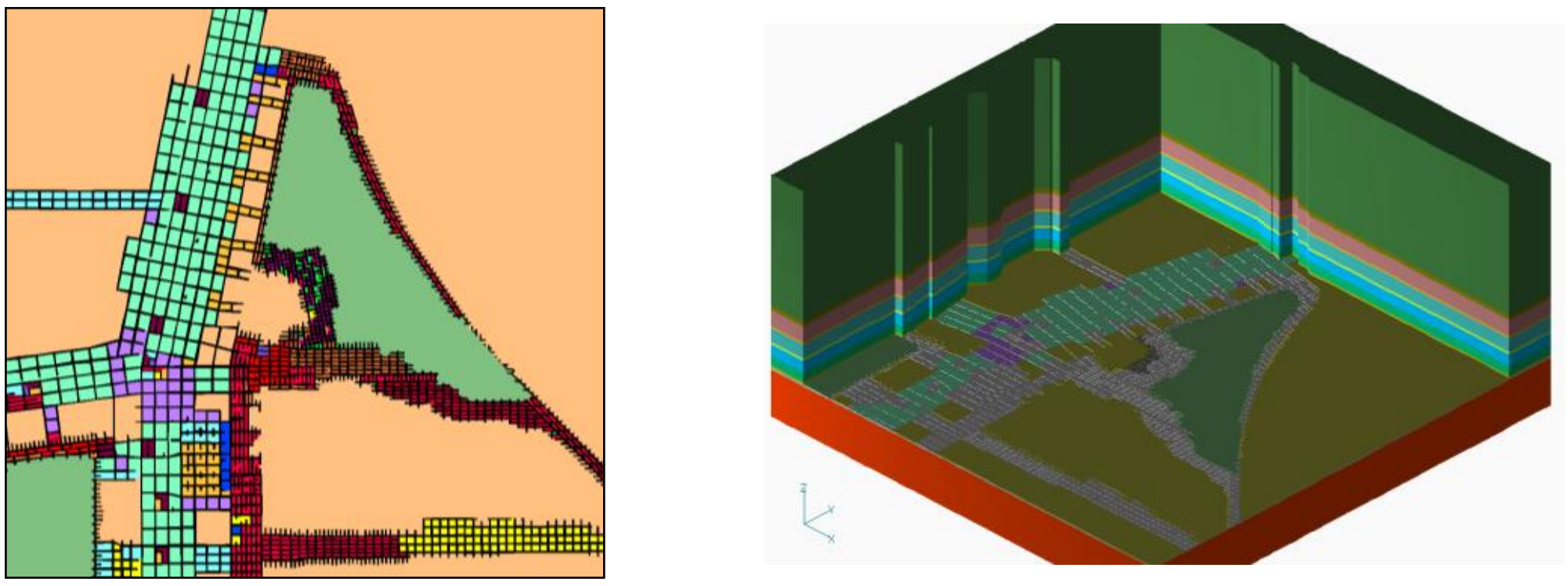

**Fig. 3. View of the 3D FEM model (right) and its horizontal cross-section (left)**

The firing of explosives in the mining face were simulated by applying hydrostatic pressure within a finite elements comprising the fired out faces in which the detonation of explosives were represented by the pressure increasing from 0 to 100 MPa within 1 ms and decreasing after detonation to zero [7]. Seismic P wave velocity was determined from seismic tomography measurements conducted at G-4/8 mining panel located within G-23 mining division of the Rudna mine (Fig. 4). The average seismic P wave velocity value equal to 5,400 m/s was obtained from tomography image presented in Figure 4 and used in numerical computations of seismic velocity seismograms in order to compare with measured seismic velocity seismograms.

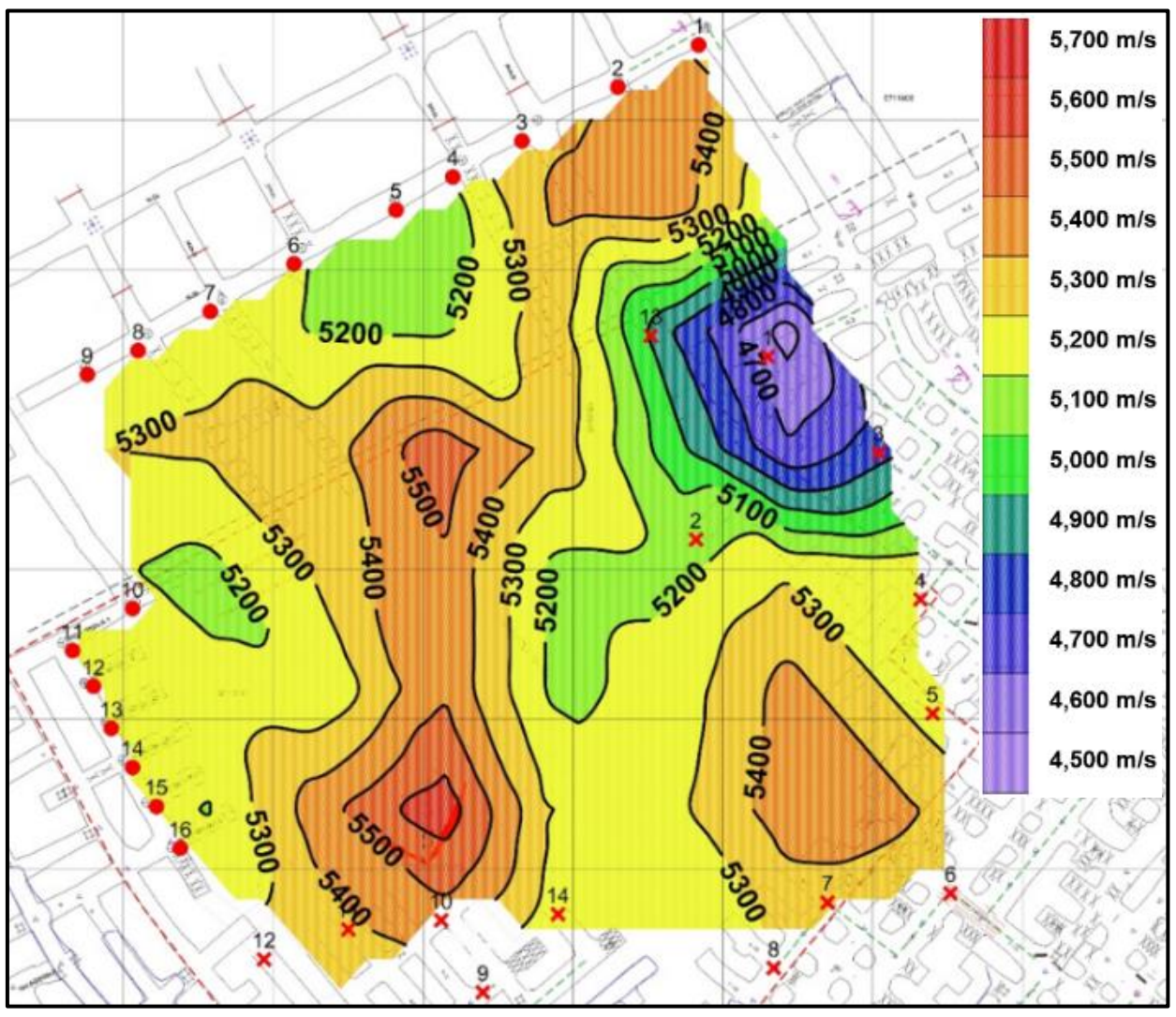

**Fig. 4. Rudna mine, Poland. G-4/8 mining panel located within G-23 mining division.**
**Tomographic velocity image (m/s) of the seismic longitudinal P wave [8]**

## 3. COMPARISON OF NUMERICAL MODEL WITH SEISMIC MEASUREMENTS RESULTS

Seismic wave packets generated by blasting in underground mines propagate in rock mass in all directions. The propagation depends on many parameters, such as structure of material, stress/strain state and the frequency of waves [9, 10]. Basically, seismic waves are subjected to scattering, reflection etc. However, those parameters were not considered in the numerical model. The peak amplitudes of recorded particle velocity values were compared with peak particle velocities and particle accelerations values obtained from numerical simulations. Though there may be noticed some differences in measured and computed values of particle velocity and particle acceleration values due to the above mentioned reasons, those differences are not significant considering all simplifications made in numerical modelling calculations. Generally, all the corresponding values are of the same order, what may confirms the appropriateness of the assumed numerical model parameters. In particular, assumptions on the rock structural damping and the time-dependent pressure type explosion model seem to be sufficiently realistic from point of view of the velocity effects at the remote distances from the firing face. Velocity and acceleration seismograms are presented in Figure 5 and 6.

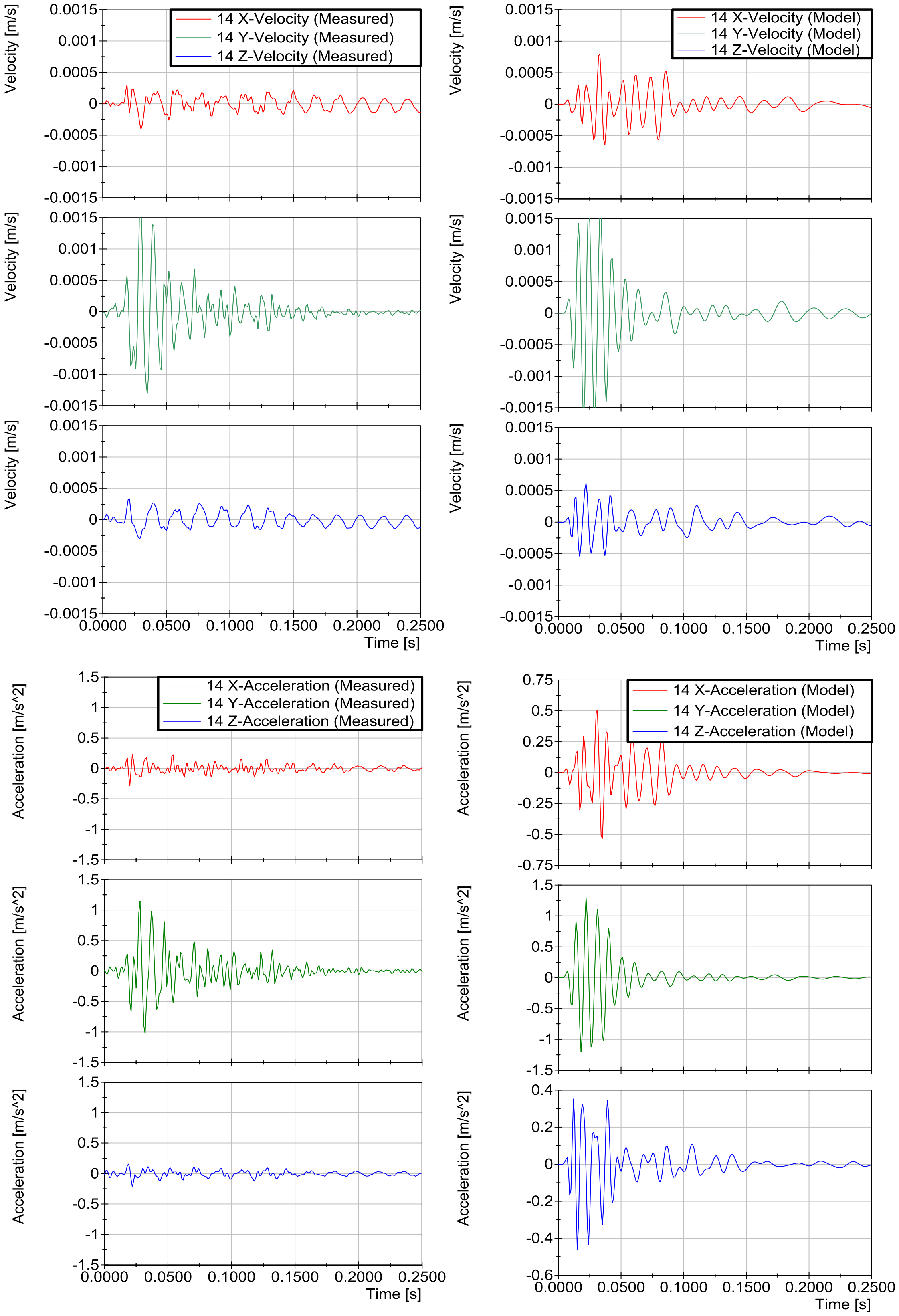

**Fig. 5. Recorded (left) and numerically calculated (right) seismic velocity and acceleration seismograms generated by production blasting: velocity (top) and acceleration (bottom) seismograms for three component geophone no. 14 shown in Figure 4**

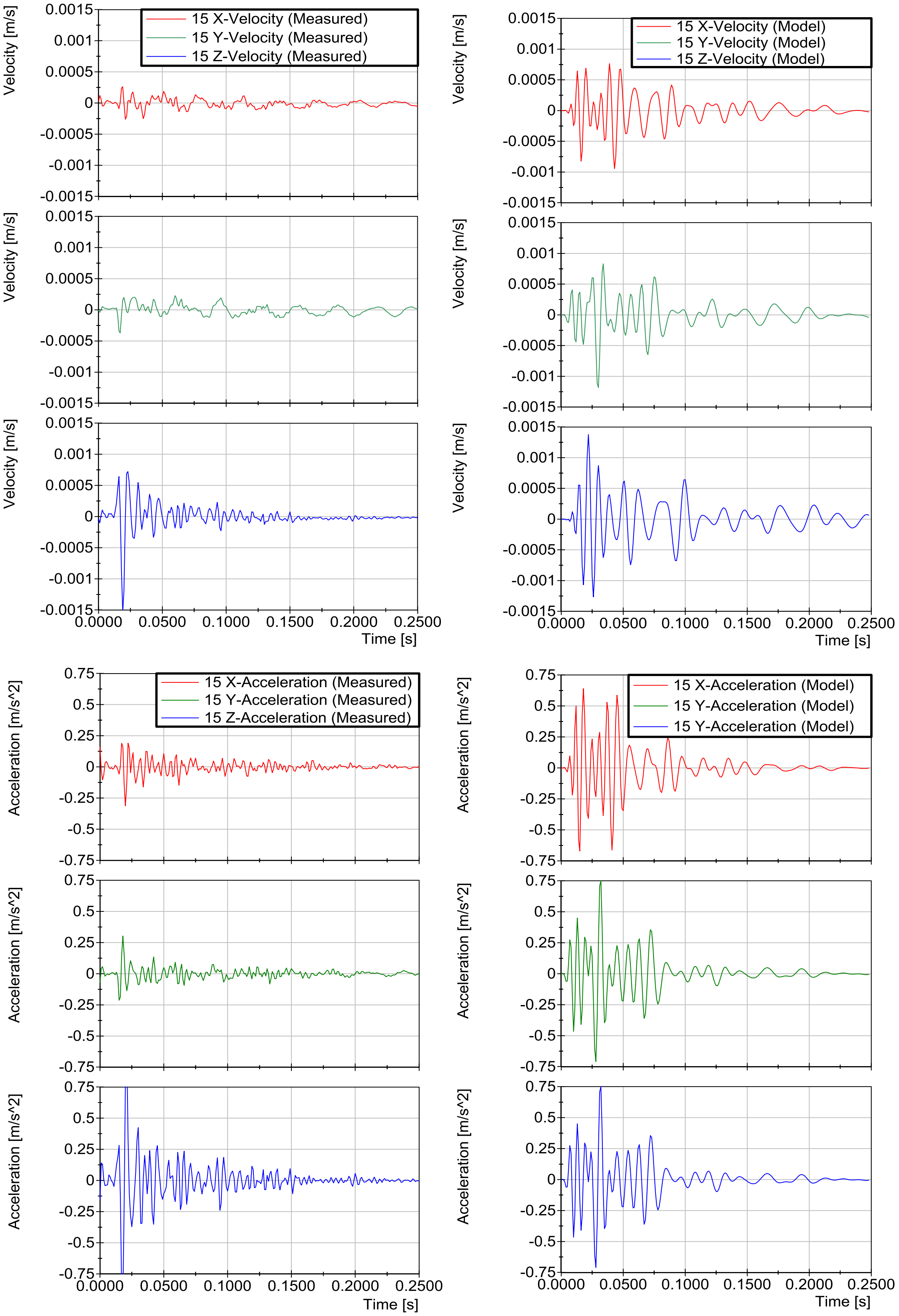

**Fig. 6. Recorded (left) and numerically calculated (right) seismic velocity and acceleration seismograms generated by production blasting: velocity (top) and acceleration (bottom) seismograms for three component geophone no. 15 shown in Figure 4**

The corresponding peak particle velocity and acceleration values, obtained by direct underground measurement and by computer calculations were compared in Figures 7 and 8. From the presented figures one may conclude that the differences between the peak particle values of velocity and acceleration determined numerical and empirical approaches differ insignificantly, proving the usefulness of the proposed numerical procedure applied for assessment of the particle velocity and acceleration computation.

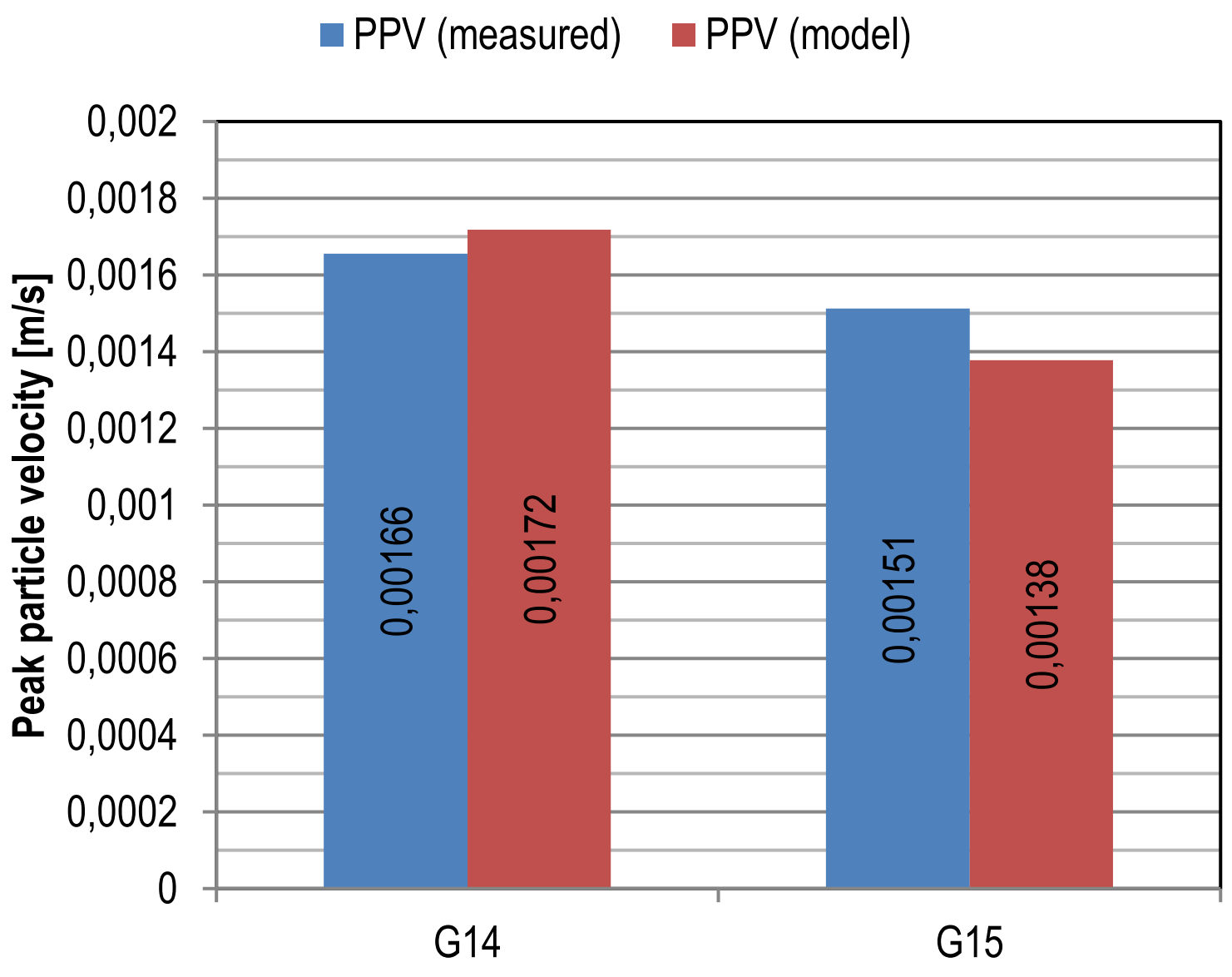

Fig. 7. Comparison of measured and numerically calculated Peak Particle Velocity values

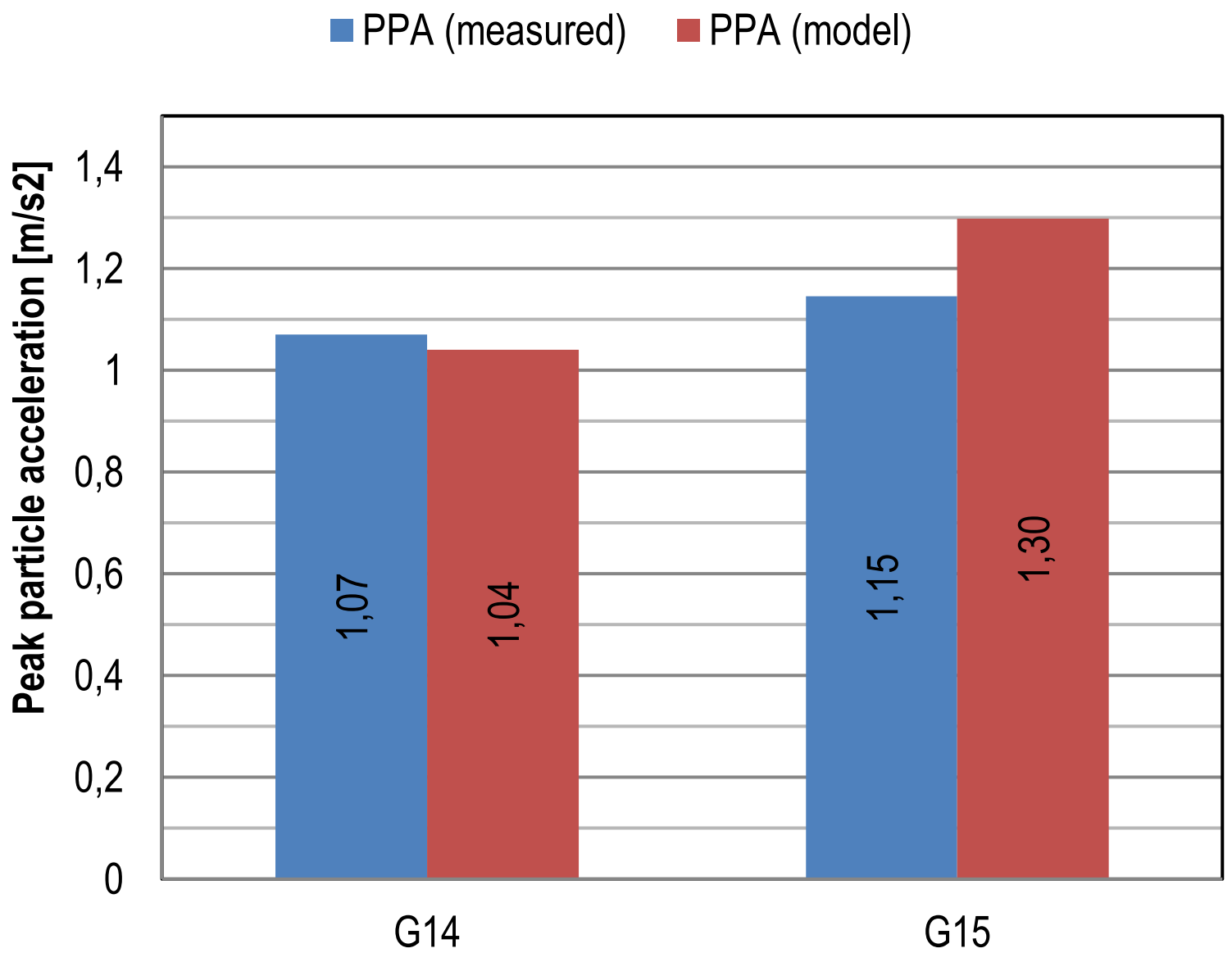

Fig. 8. Comparison of recorded and numerically calculated Peak Particle Acceleration values

## 4. CONCLUSIONS

The primary objective of this paper was to develop an appropriate method for numerical simulation of the seismic wave in order to provide an effective tool for assessment of the seismic effect of blasting works, what in turn should result in improvement of the roof control methods of deep underground workings.

Based on investigations of seismic wave analysis generated by detonation of explosives in a specific location of the rock mass from both measured and theoretical point of view one may conclude, that appropriate selection of the source of vibration (amplitude, frequency etc.) spatiotemporal distribution of seismic particle velocities can be simulated using numerical tools with acceptable accuracy. Particle velocity values calculated numerically are coherent with those recorded in situ.

## ACKNOWLEDGEMENTS

This paper has been prepared through the Horizon 2020 EU funded project on “Sustainable Intelligent Mining Systems (SIMS)”, Grant Agreement No. 730302.